\documentclass[aps,prl,twocolumn,groupedaddress,superscriptaddress,preprintnumbers]{revtex4-1}
\usepackage{amsmath}
\usepackage{amssymb}
\usepackage{graphicx}
\usepackage{hyperref}
\usepackage{xspace,slashed}
\usepackage[utf8]{inputenc}

\newcommand{\CERNaff}{Theoretical Physics Department, CERN, CH-1211, Geneva 23, Switzerland} 
\newcommand{\LisbonAff}{LIP, Av. Prof. Gama Pinto, 2, 1649-003 Lisboa, Portugal}

\begin{document}

\title{Sensitivity of jet substructure to jet-induced medium response}

\author{Jos\'{e} Guilherme Milhano}
\affiliation{\LisbonAff}
\affiliation{\CERNaff}
\author{Urs Achim Wiedemann}
\affiliation{\CERNaff}
\author{Korinna Christine Zapp}
\affiliation{\LisbonAff}
\affiliation{\CERNaff}

\begin{abstract}
Jet quenching in heavy ion collisions is expected to be accompanied by recoil effects, but unambiguous signals
for the induced medium response have been difficult to identify so far. Here, we argue that modern jet substructure
measurements can improve this situation qualitatively since they are sensitive to the momentum distribution
inside the jet. We show that the groomed subjet shared momentum fraction $z_g$, and the girth
of leading and subleading subjets signal recoil effects with dependencies that are
absent in a recoilless baseline. We find that recoil effects can explain most of the medium modifications 
to the $z_g$ distribution observed in data. Furthermore, for jets passing the Soft Drop Condition, recoil effects induce 
in the differential distribution of subjet separation $\Delta R_{12}$ a characteristic increase with $\Delta R_{12}$,
and they introduce a characteristic enhancement of the girth of the subleading subjet with decreasing $z_g$. 
We explain why these qualitatively novel features, that we establish in \textsc{Jewel+Pythia} simulations, reflect
generic physical properties of recoil effects 
that should therefore be searched for as telltale signatures of jet-induced medium response.
\end{abstract}

\pacs{12.38.Mh,  13.87.-a, 25.75.-q, 25.75.Bh, 12.38.-t, 24.85.+p}
\preprint{CERN-TH-2017-150, MCnet-17-12}

\maketitle 

High momentum transfer processes with hadronic final states are generically and strongly modified when occuring
within the dense environment produced in nucleus-nucleus collisions. This jet quenching phenomenon 
is being studied systematically at the LHC for jet $p_\perp$-spectra, dijet asymmetries, jet fragmentation functions, jet shapes and, 
most recently,  for a large class of increasingly refined jet substructure observables. 
Jet quenching implies jet-medium interactions. If the medium is close to a perfect liquid, medium recoil propagates 
in the form of hydrodynamic excitations~\cite{CasalderreySolana:2004qm}, but it is expected to show signs of large angle 
scattering if jet-medium interactions were to resolve partonic degrees of freedom in the medium~\cite{Zapp:2008gi,DEramo:2010wup,Kurkela:2014tla}. Beyond confirming the assumed dynamics of jet-medium
interactions, the observation of recoil distributions is thus of 
great interest for characterizing the nature of the medium. 

However, the characterization of jet recoil distributions has remained elusive so far for several reasons.
In particular, recoil effects are expected 
to contribute mainly to the soft large-angle hadronic activity, but there are experimental and theoretical uncertainties in establishing soft recoil remnants on top of a large and fluctuating background that need to be controlled. 
Also, many of the measurements used to characterize jet quenching are remarkably insensitive to soft large-angle activity. 
For instance, quenched hadron spectra are by construction insensitive to how the lost energy is distributed, and traditional
jet quenching observables constructed from jet $p_\perp$ and  jet axis (such as the jet nuclear modification factor)
are dominated by hard contributions. One may expect that jet shape observables are more sensitive to the jet 
medium response since they are sensitive to momentum distributions inside the jet. 
Our main result
will be to establish a first example for a combination of jet substructure observables --- namely measurements of the subjet 
shared momentum fraction $z_g$ and of girth --- that allow for the
separation of recoil effects from alternative interpretations with both characteristic quantitative and qualitative
features in the data.

In contrast to jet quenching models that parametrize (e.g., in terms of  the
quenching parameters $\hat{q}$  and $\hat{e}$) the recoil carried away from the jet, fully dynamical event generators of jet quenching 
are better suited to study recoil effects as they can propagate them into final state particle distributions. 
To exploit the resulting phenomenological opportunities, however, one needs a robust prescription for separating 
medium recoils from the initial thermal component of recoiling partons that is part of the soft background activity. 
For the event generator \textsc{Jewel}~\cite{Zapp:2013vla}, such a tool was validated recently in 
Ref.~\cite{KunnawalkamElayavalli:2017hxo}. 
It enables us in the present study to distinguish in fully dynamical jet quenching simulations between effects that 
are due to the splitting of jet constituents, and effects that arise from momentum transfers to recoiling medium scattering 
centers. We emphasize that while both effects are part of the same physical process, there is no model parameter that 
would allow one to vary their relative strength and trade one for the other \cite{foot1}.
 Moreover, both effects manifest themselves 
differently in different kinematical regions. It is in this sense that we can separate here both effects operationally in a 
model-indepent way. 

\textsc{Jewel} has been tested against a large class of jet quenching measurements, including traditional jet observables built from the
jet $p_\perp$ and axis (such as jet $R_{AA}$, dijet asymmetry $A_J$)~\cite{Milhano:2015mng,Zapp:2012ak}, as well 
as jet shape observables that are more sensitive to medium effects~\cite{KunnawalkamElayavalli:2017hxo}. The simulations shown in this work are based 
on di-jet events generated in the standard setup~\cite{Zapp:2013vla} at $\sqrt{s_{\rm NN}}=5.02$ TeV. No attempt was
made to improve comparison to data by retuning model parameters. While the following discussion focuses
on the physics of two particular jet substructure observables, we emphasize that \textsc{Jewel} with the model parameter 
settings used here is documented to provide a correct qualitative and good quantitative description of jet quenching in general.

The Soft Drop algorithm~\cite{Dasgupta:2013ihk,Larkoski:2014wba} reconstructs jets with the anti-$k_\perp$ algorithm~\cite{Cacciari:2008gp} and reclusters them with a prescription entirely based on angles (Cambridge/Aachen). 
The last step of this reclustering is then undone to give the two prongs with the largest angular separation. 
If the $p_\perp$-sharing between the two prongs satisfies
\begin{equation}
	z_g \equiv \frac{{\rm min}\left( p_{\perp,1},p_{\perp,2}\right)}{p_{\perp,1} + p_{\perp,2}} > z_{\rm cut} 
	\left( \frac{\Delta R_{12}}{R}\right)^\beta\, ,
	\label{eq1}
\end{equation}
then the prongs are accepted and the algorithm terminates. Otherwise, the softer of the prongs is rejected, the last
reclustering step on the hard prong is undone, and the algorithm continues till condition (\ref{eq1}) is satisfied. 
This is one of a variety of grooming techniques that can be used to systematically reject (or study) 
soft contributions associated to jets. In eq.~(\ref{eq1}), $R$ denotes the jet radius. In the following, we work for 
$\beta = 0$, and we use the default $z_{\rm cut}=0.1$. We also require that only configurations with $\Delta R_{12} > 0.1$
are included in the $z_g$-distribution. This condition was added by the CMS collaboration to the original Soft Drop proposal, 
and we adopt it to facilitate comparison to the preliminary data~\cite{CMS:2016jys}. 

 Here, we investigate the physical mechanisms underlying the softening of the groomed shared momentum fraction $z_g$
 in \textsc{Jewel}, including the possibility that recoil effects contribute. In general, the momentum of recoiling partons is composed
 of a thermal component that they carry before the jet-medium interaction, as well as the momentum transferred when
 interacting with jet constituents. Only the latter contributes to the medium response, the former is removed experimentally
 by background subtraction techniques. However, these techniques cannot be applied to \textsc{Jewel} as it 
 does not generate full heavy ion events.  Instead, consistent with experimental procedures,
 the (thermal) background contribution is subtracted from generated event samples with a so-called 4-momentum subtraction
 technique validated in~\cite{KunnawalkamElayavalli:2017hxo}. 
 
\begin{figure}
\begin{center}
\includegraphics[width=0.95\linewidth]{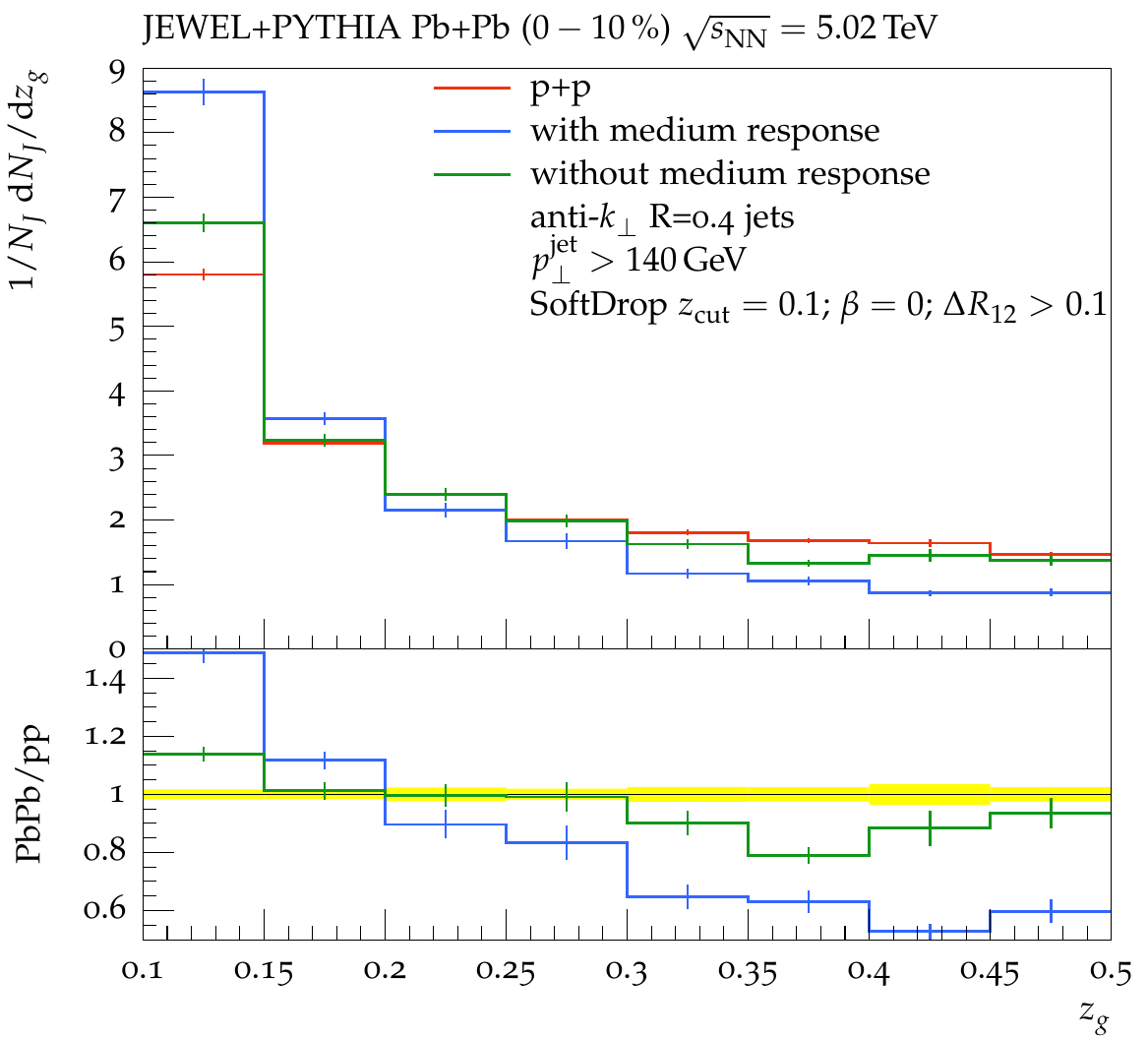}
\includegraphics[width=0.95\linewidth]{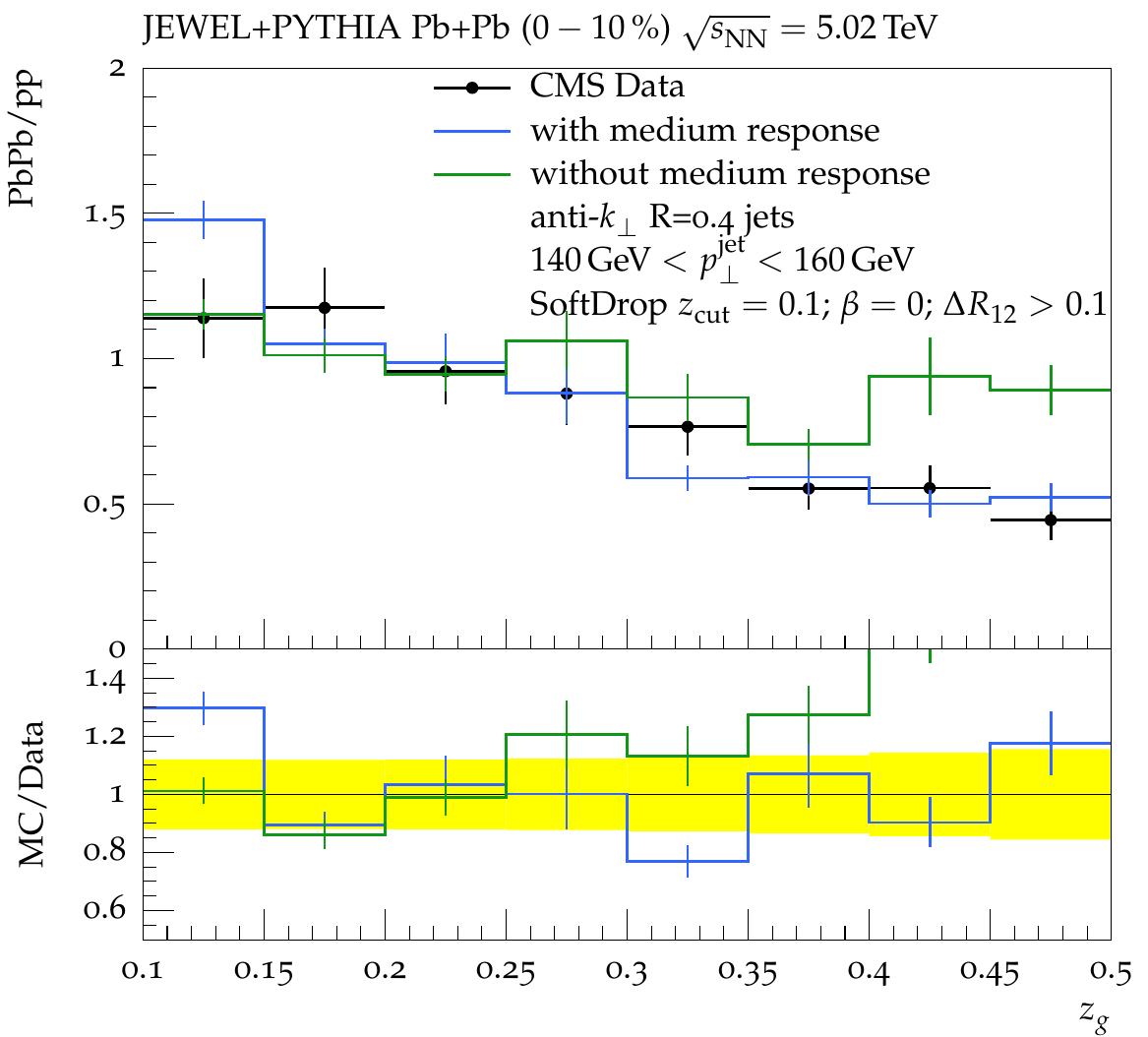}
\end{center}
\caption{(top) \textsc{Jewel+Pythia} result for the groomed shared momentum fraction $z_g$ in central PbPb events 
analyzed with (blue curve) and without (green curve) keeping track of medium response and compared to simulated pp events
(red curve). (bottom) The ratio of the $z_g$-distributions in PbPb and pp events, compared to CMS data for 
jet $p_\perp$ between 140 GeV and 160 GeV. 
All results are for $\sqrt{s_{\rm NN}} = 5.02$ TeV and are shown background subtracted (4-momentum subtraction method) 
and on hadron level. 
}
\label{fig1}
\end{figure}

 We emphasize that for hadronization, \textsc{Jewel} 
 converts
 all recoiling partons into gluons that are inserted 
 into the strings that connect the partons forming the jets. It is therefore not meaningful to label hadrons in the event record
 as belonging to the jet or to the medium response. However, one can hadronize events in \textsc{Jewel} with or without the recoiling partons. Fig.~\ref{fig1} shows the corresponding $z_g$-distributions. Since recoiling partons do not rescatter in \textsc{Jewel}, and since 
 rescattering induces thermalization processes, generated events with recoiling  partons may overestimate the physically 
 expected medium response. The truth is therefore expected to lie in between the green (without recoil) and blue 
 (with recoil) curves in Fig.~\ref{fig1}, and the difference between both curves should be regarded as an upper bound for 
 the expected medium-response. 
 %
%%%%%%%%%%%%%%%%%%%%%%%%%%%%%%%%%
 %%%%%%%%%%%%%%%%%%%%%%%%%%%%%%%%%
 
 Even without including recoiling partons, the simulated $z_g$-distribution in Fig.~\ref{fig1} shows a mild tilt towards smaller 
 $z_g$ in comparison to the proton-proton baseline. Without additional information, the interpretation of this tilt remains
 ambiguous. The reason is that the $z_g$-distribution is a  self-normalizing curve. A tilt of the type shown in 
 Fig.~\ref{fig1} can therefore arise either (i) from an enhanced contribution at small $z_g$ (that reduces the bin entries at 
 large $z_g$ due to normalization), or (ii) from a depletion of jets with large $z_g$ (that would
 enhance bin entries at small $z_g$ by normalization). The first of these two possibilities has been 
 argued~\cite{Chien:2016led,Mehtar-Tani:2016aco} to be the dominant one, based on the following two observations:
 first, to lowest perturbative order in QCD (and without medium-effects), the $z_g$-distribution $p(z_g)$ 
 for $\beta = 0$ is given by the LO QCD splitting functions $P(z)$ ~\cite{Larkoski:2015lea}
\begin{equation}
	p(z_g) = \frac{P\left(z_g\right) + P\left(1-z_g\right)}{\int_{z_{\rm cut}}^{1/2} dz \left[ P\left(z_g\right) + P\left(1-z_g\right) \right]}\, ,
	\label{eq2}
\end{equation}
and second, medium-induced gluon radiation is expected to soften the perturbative splitting functions. Therefore, if one neglects
recoiling partons, the medium-induced enhancement of gluon splittees in the parton shower provides a candidate mechanism 
for enhancing the fraction of subleading subjets with small groomed momentum fraction $z_g$. However, for this mechanism to
be efficient, medium-induced gluon radiation must be sufficiently hard to pass the cut (\ref{eq1}). Inspection of generated 
events reveals that this condition is rarely satisfied in \textsc{Jewel}. Indeed, while medium-induced parton splitting underlies
the simulation of jet quenching in \textsc{Jewel}, partonic splittees induced by jet-medium interactions carry rarely a sufficient
energy $O\left(E_{\rm jet}\, z_g\right)$ to make it above the cut (\ref{eq1}), and hadronization reduces this contribution further.
Also, in simulations without recoiling partons, the likelihood of medium-induced splittees to cluster with other jet fragments 
to subjets that pass the cut (\ref{eq1}) is small. Rather the dominant contribution to the small tilt of 
$(1/N_J)\mathrm{d}N_J/\mathrm{d}z_g$ in simulations without recoiling partons comes from 
the fact that all partons in the shower undergo parton energy loss and that this suppresses in particular the
yield of events with large $z_g$. As jets with a large $z_g$ will show a softer fragmentation, 
this is  consistent with earlier observations that such broader jets are more susceptible to energy loss and thus more likely
to fail analysis cuts~\cite{Milhano:2015mng,Rajagopal:2016uip,Casalderrey-Solana:2016jvj}. We have checked 
this statement for the present analysis (data not shown).
%
%%%%%%%%%%%%%%%%%%%%%%%%%%%%%%%%%
\begin{figure}
%\hspace*{40mm}
\begin{center}
\includegraphics[width=0.95\linewidth]{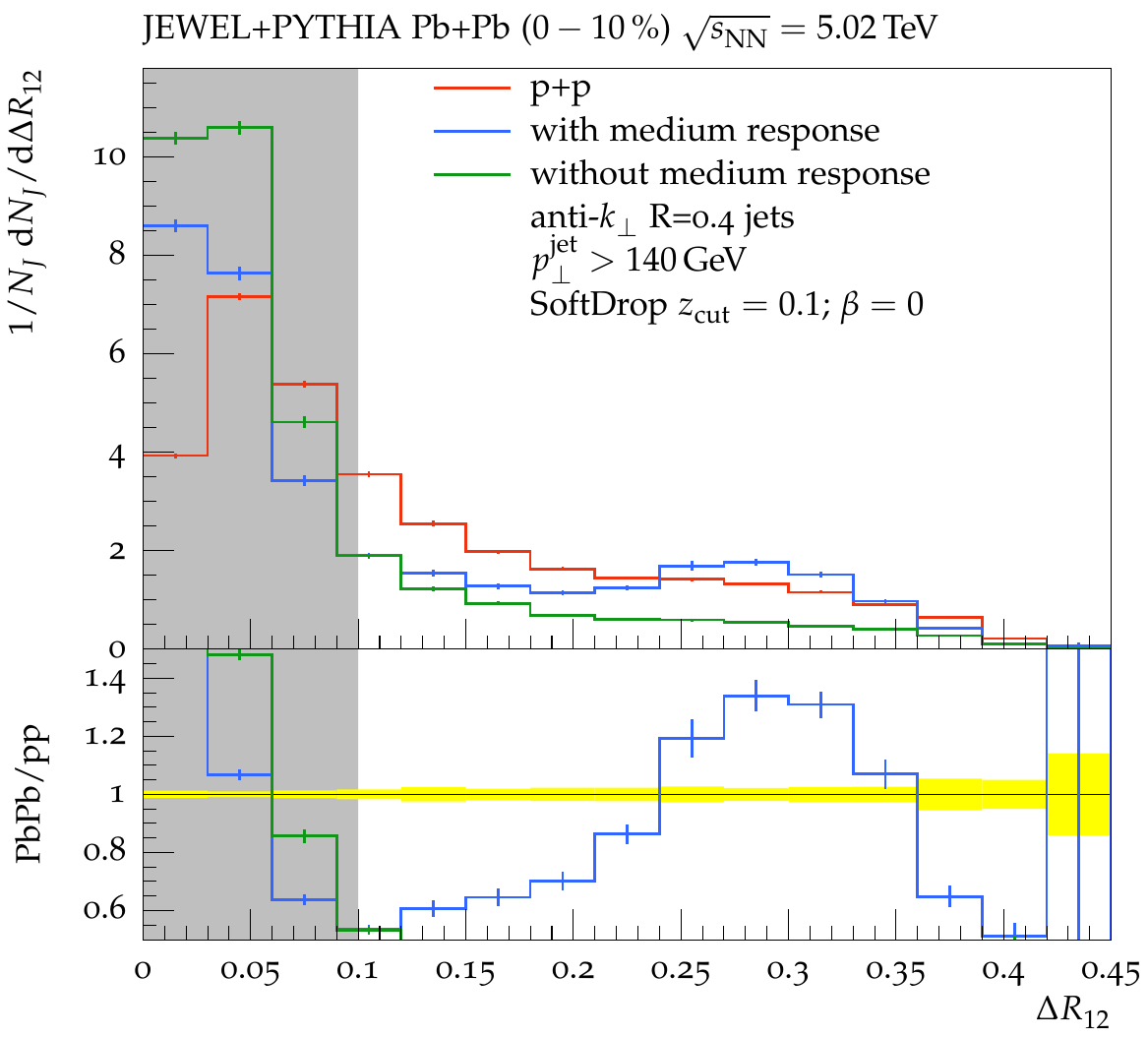}
\end{center}
\caption{Distribution in the relative separation $\Delta R_{12}$ of the two subjets for jets that pass the Soft Drop condition (\ref{eq1}),
supplemented by the $\Delta R_{12}>0.1$ requirement (grey band). 
}
\label{fig2}
\end{figure}
 %%%%%%%%%%%%%%%%%%%%%%%%%%%%%%%%%
 
 Once recoiling partons are included in the analysis, the tilt in the $z_g$-distribution increases significantly and the shape
 is in quantitative agreement with experimental data (see r.h.s. of Fig.~\ref{fig1}). In contrast to the case without recoil, the 
 dominant contribution to the tilt comes now from an enhancement of jets with soft subleading subjets that pass the grooming 
 cut (\ref{eq1}). The reason is that soft large-angle recoil contributions get clustered into (sub)jets and can thus promote candidate 
 prongs of low $z$ to above the Soft Drop condition (\ref{eq1}). Our simulations thus
 suggest that the long-sought medium response that provides a negligible or difficult to discriminate contribution in many 
 other jet quenching observables may dominate the $z_g$ distribution. We next ask to what extent this interpretation can 
 be corroborated by complementary measurements.
 
 To this end, we study first for the jet sample that contributes to the $z_g$-distribution the relative separation $\Delta R_{12}$ 
 in the $\Delta\eta \times \Delta\phi$-plane between the leading and subleading prongs. As described above, jets with broader
 fragmentation patterns are expected to fail analysis cuts such as (\ref{eq1}) more easily. Consistent with this picture, in the 
 absence of recoil effects (see green curve on the r.h.s. of Fig.~\ref{fig2})
 the fraction of jets with large $\Delta R_{12}$ that pass the analysis cut is strongly reduced.
 If medium response is included in the analysis, the $\Delta R_{12}$-distribution changes qualitatively in a very characteristic
 way. The reason is that if a subleading candidate prong is further separated from the leading prong, then there is a larger 
 area in the $\Delta\eta \times \Delta\phi$-plane from which soft recoil contributions can be clustered together with this soft
 prong. This makes it more likely to promote soft prongs above the Soft Drop condition (\ref{eq1}) if $\Delta R_{12}$ is
 larger. As a consequence, the $\Delta R_{12}$-distribution
 increases with increasing  $\Delta R_{12}$ up to a separation scale that is set by the jet radius. Therefore, the $\Delta R_{12}$-distribution (blue curve) peaks at
 a value $\Delta R_{12}$ somewhat smaller than $R$.  We conclude that the increase of the  $\Delta R_{12}$-distribution 
 with increasing $\Delta R_{12}$ would be a characteristic telltale sign for the dominance of recoil effects in 
 medium-modifications of the groomed shared momentum fraction $z_g$. 
%
%%%%%%%%%%%%%%%%%%%%%%%%%%%%%%%%%
\begin{figure}
\begin{center}
\includegraphics[width=0.95\linewidth]{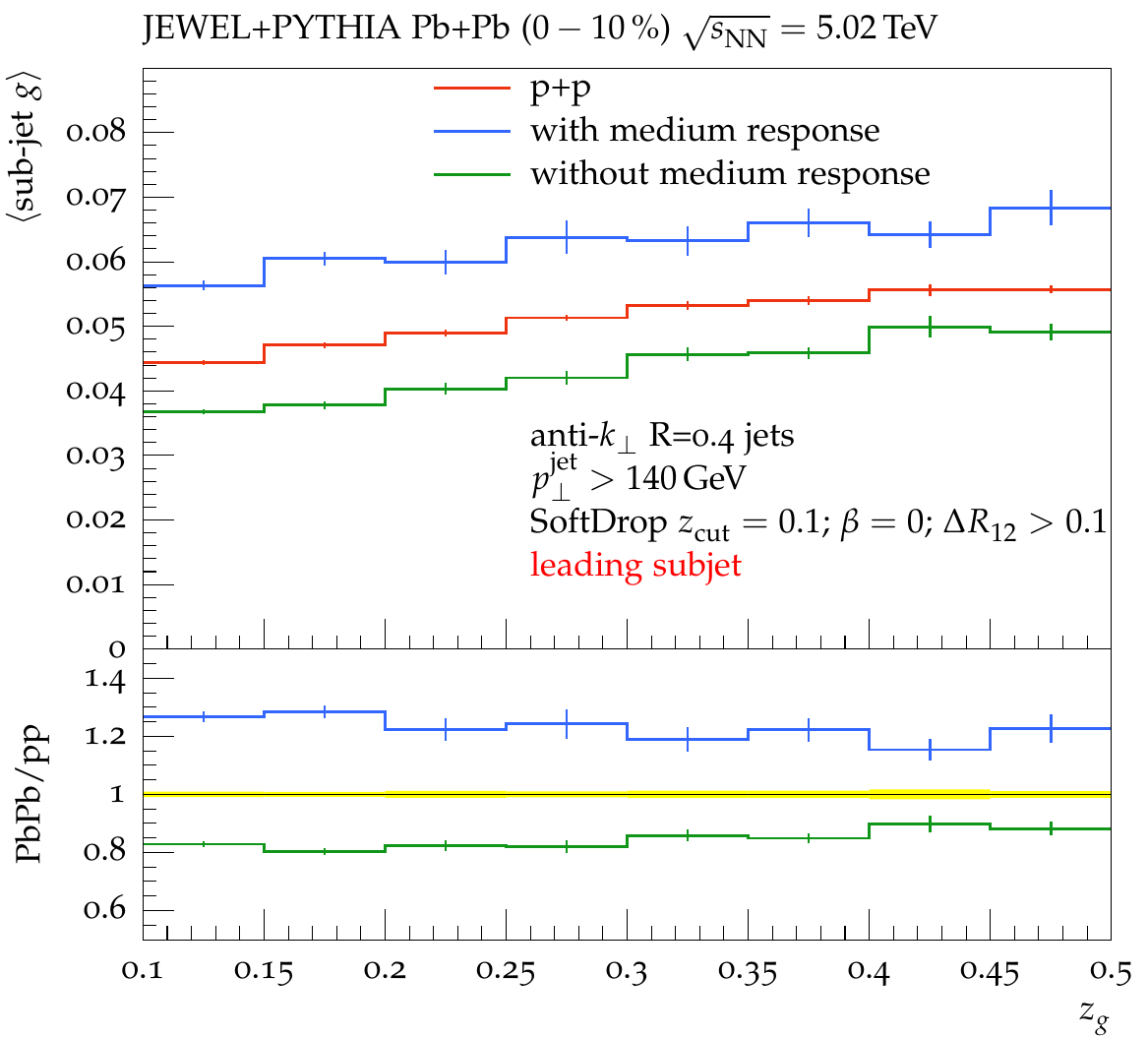}
\includegraphics[width=0.95\linewidth]{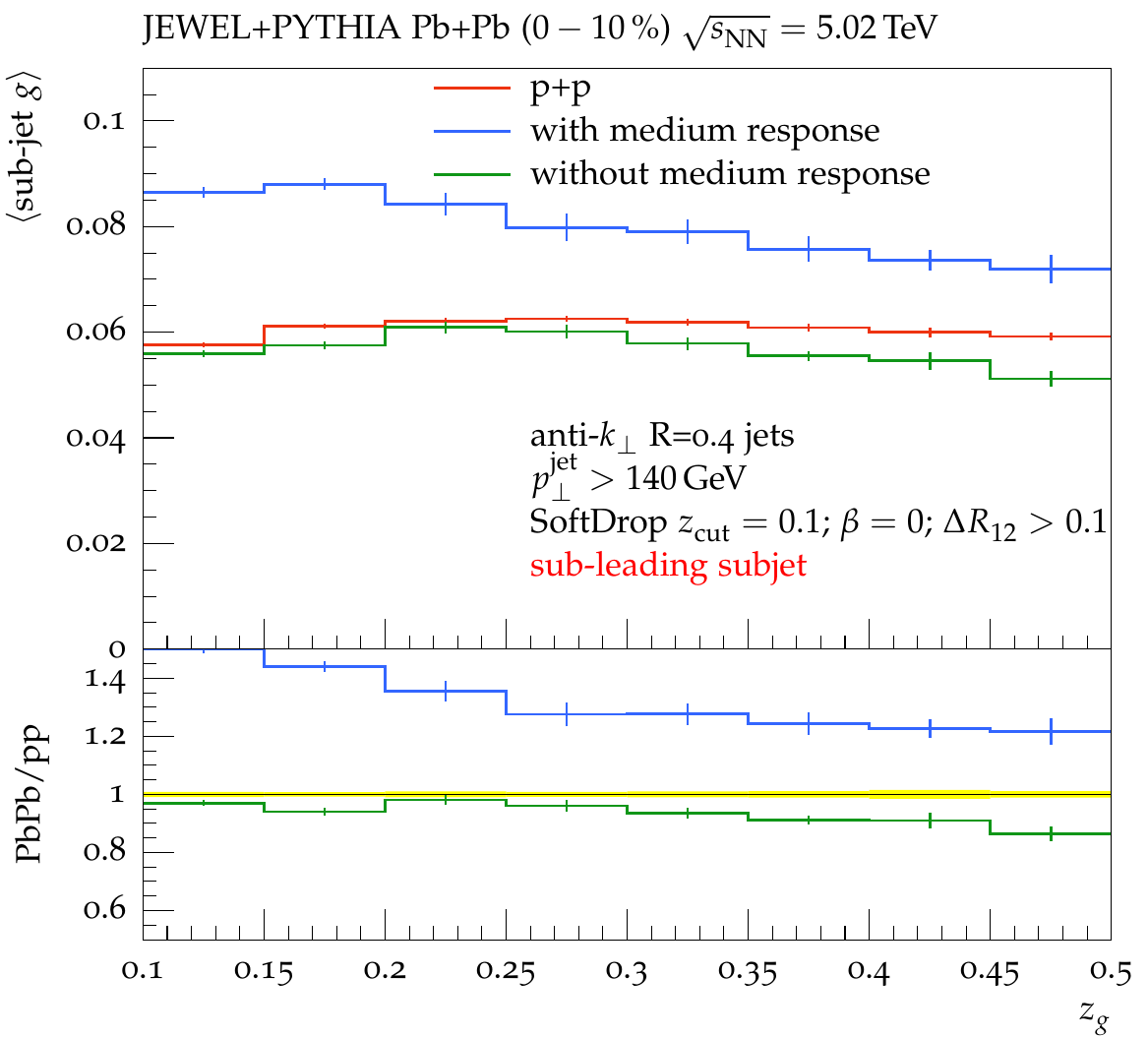}
\end{center}
\caption{Jewel+Pythia results for the first radial moment (\ref{eq2}) (girth $g$)  of the leading (l.h.s.) and
subleading (r.h.s.) subjet in jets reconstructed with the anti-$k_\perp$ algorithm for $R=0.4$  and 
that pass the Soft Drop condition (\ref{eq1}). Results are for jets simulated on hadron level in
$\sqrt{s_{\rm NN}} = 5.02$ TeV PbPb collisions  with (blue curve) and without (green curve) recoil effects,
as well as for proton-proton reference data (red curve). The 4-momentum subtraction method is used to 
provide background subtracted data consistent with experimental procedures.  
}
\label{fig3}
\end{figure}
%%%%%%%%%%%%%%%%%%%%%%%%%%%%%%%%%%% 
 
 By now, several independent model studies support the at least partial cancellation of two qualitatively
 different effects in many jet quenching observables~\cite{Casalderrey-Solana:2016jvj,Tachibana:2017syd,KunnawalkamElayavalli:2017hxo}. On the one hand, parton energy loss effectively peels off soft components from the jet, thereby narrowing the jet core. On the other hand, 
 medium response can counteract this tendency as recoil effects contribute to jet broadening. The interplay of both 
 effects has been observed to be at work also in some jet shape observables, including  jet mass and girth~\cite{Casalderrey-Solana:2016jvj,KunnawalkamElayavalli:2017hxo}. However, the kinematical
 distribution of recoil is generally different from that of medium-induced radiation, and despite partial cancellation of
 both effects, differential distributions in other jet substructure observables may thus be expected to maintain some characteristic
 sensitivity to medium response. Here, we discuss this possibility for girth $g$, which is defined by summing over the
 momenta $p_\perp^{(k)}$ of all consituents of the jet with a weight given by the distance $ \Delta R_{kJ}$ from the jet 
 axis, 
 \begin{equation}
 	g = \frac{1}{p_\perp^{\rm jet}} \sum_{k\in J} p_\perp^{(k)}\, \Delta R_{kJ}\, .
 	\label{eq2}
 \end{equation}
 In general, this radial moment of the jet profile is expected to increase with recoil effects that broaden the jet,
 and it is expected to decrease if radiation narrows the jet core by peeling off preferentially soft large angle components.   Both
 mechanisms are clearly seen at work for the girth of the leading subjet, where the girth in PbPb events reconstructed
 without recoil effects is seen to be reduced compared to the pp baseline, while the girth is increased in events 
 including recoil effects (top panel, Fig.~\ref{fig3}). Both effects cancel partially, consistent with earlier observations. 
 For the leading subjet, the net effect is a shift of the magnitude of girth that is approximately independent of $z_g$.
 
 The situation is somewhat different for the girth of the subleading subjet. First, in the absence of recoil effects, jet
 quenching leads to a much smaller reduction of girth. The reason is that the jet can only get narrower 
 by losing energy, but subleading subjets cannot lose much energy without failing the Soft Drop condition (\ref{eq1}). 
 If the subleading subjet fraction $z_g$ is larger, then this bias
 is less significant, and this explains the slight increase in the reduction of girth with increasing $z_g$.
 On the other hand, for the case in which recoil effects are included in the analysis, the girth of subleading subjets is 
 approximately a factor 2 more strongly enhanced for small $z_g \simeq 0.1$ than for $z_g \simeq 0.5$. This is
 an independent test of the argument that the tilt of the $z_g$-distribution is mainly due to recoil effects that
 promote soft candidate prongs above the Soft Drop cut condition (\ref{eq1}):  if subleading subjets
 at low $z_g$ have a pronounced recoil contribution, then they are expected to be particularly broad, and this
 is what is reflected in a more strongly enhanced girth at small $z_g$. The combined analysis of the girth of leading
 and subleading subjets provides thus independent sensitivity to recoil  effects  and can therefore help to disentangle 
 effects from medium response in jet quenching models. 
 
 We finally dare to share our experience that physics conclusions about the presence of recoil effects can only be drawn 
 from models of a certain technical maturity. For instance, hadronization effects are also known to contribute to the
 broadening of jets. In our simulations, the girth one extracts from generated data at (unobservable) parton level shows 
 qualitatively similar but much stronger recoil effects than the data on hadron level discussed here. The use of an
 independently validated hadronization prescription is therefore important for arriving at realistic physics conclusions. 
 An analogous remark applies to the use of background subtraction techniques. 
 
 The $z_g$-distribution and the girth of subjets are not the only jet measurements that are sensitive to recoil effects.
 Recent studies indicate that also the ratio of jet fragmentation functions, the jet mass and the radial jet profile show
 characteristic dependencies that are naturally accounted for by recoil effects~\cite{KunnawalkamElayavalli:2017hxo,Tachibana:2017syd}.
 Recoil effects have also been argued to affect $\gamma$-hadron azimuthal correlations~\cite{Chen:2017zte} and single 
 inclusive jet measurements~\cite{Wang:2016fds}.
 We expect that a more complete analysis of medium response in jet quenching will profit from the totality of
 modern jet and in particular jet substructure measurements. In the present work, we have shown for the cases of $z_g$-distribution
 and girth that recoil effects found in a detailed simulation can be understood in terms of generic physical properties
 of recoil effects. Jet substructure measurements are sufficiently differential to test a recoil interpretation in
 multiple complementary ways. In particular, beyond demonstrating that the preliminary data for the tilt in the
 $z_g$-distributions are consistent with a recoil interpretation, we have argued that the same interpretation implies 
 characteristic features in the $\Delta R_{12}$-distribution (increase with increasing $\Delta R_{12}$) and in the girth 
 (increased enhancement of the girth of the subleading subjet at small $z_g$). These developments make us 
 confident that 15 years after the first experimental indications for jet quenching, we are in a position to constrain
 the so far elusive but logically unavoidable counterpart of jet quenching: the jet-induced medium response.   
 
\begin{acknowledgments}
We thank L. Apolin\'ario, K. Kauder, Y. Mehtar-Tani, J. Thaler, K. Tywoniuk, and M. Verweij for helpful discussions. The work of JGM and KCZ was supported by Funda\c c\~ao para a Ci\^encia e a Tecnologia (Portugal) under project CERN/FIS-NUC/0049/2015 and contracts `Investigador FCT - Development Grant' IF/00563/2012 (JGM) and `Investigador FCT - Starting Grant' IF/00634/2015 (KCZ).
\end{acknowledgments}

\bibliographystyle{elsarticle-num}

\begin{thebibliography}{00}

%\cite{CasalderreySolana:2004qm}
\bibitem{CasalderreySolana:2004qm}
  J.~Casalderrey-Solana, E.~V.~Shuryak and D.~Teaney,
  %``Conical flow induced by quenched QCD jets,''
  J.\ Phys.\ Conf.\ Ser.\  {\bf 27} (2005) 22
   [Nucl.\ Phys.\ A {\bf 774} (2006) 577]
  doi:10.1016/j.nuclphysa.2006.06.091, 10.1088/1742-6596/27/1/003
  [hep-ph/0411315].
  %%CITATION = doi:10.1016/j.nuclphysa.2006.06.091, 10.1088/1742-6596/27/1/003;%%
  %364 citations counted in INSPIRE as of 02 May 2017

%\cite{Zapp:2008gi}
\bibitem{Zapp:2008gi}
  K.~Zapp, G.~Ingelman, J.~Rathsman, J.~Stachel and U.~A.~Wiedemann,
  %``A Monte Carlo Model for 'Jet Quenching',''
  Eur.\ Phys.\ J.\ C {\bf 60} (2009) 617
  doi:10.1140/epjc/s10052-009-0941-2
  [arXiv:0804.3568 [hep-ph]].
  %%CITATION = doi:10.1140/epjc/s10052-009-0941-2;%%
  %117 citations counted in INSPIRE as of 02 May 2017
  
  %\cite{DEramo:2010wup}
\bibitem{DEramo:2010wup}
  F.~D'Eramo, H.~Liu and K.~Rajagopal,
  %``Transverse Momentum Broadening and the Jet Quenching Parameter, Redux,''
  Phys.\ Rev.\ D {\bf 84} (2011) 065015
  doi:10.1103/PhysRevD.84.065015
  [arXiv:1006.1367 [hep-ph]].
  %%CITATION = doi:10.1103/PhysRevD.84.065015;%%
  %90 citations counted in INSPIRE as of 02 May 2017
  
  %\cite{Kurkela:2014tla}
\bibitem{Kurkela:2014tla}
  A.~Kurkela and U.~A.~Wiedemann,
  %``Picturing perturbative parton cascades in QCD matter,''
  Phys.\ Lett.\ B {\bf 740} (2015) 172
  doi:10.1016/j.physletb.2014.11.054
  [arXiv:1407.0293 [hep-ph]].
  %%CITATION = doi:10.1016/j.physletb.2014.11.054;%%
  %28 citations counted in INSPIRE as of 02 May 2017

 %\cite{Zapp:2013vla}
\bibitem{Zapp:2013vla}
  K.~C.~Zapp,
  %``\textsc{Jewel} 2.0.0: directions for use,''
  Eur.\ Phys.\ J.\ C {\bf 74} (2014) no.2,  2762
  doi:10.1140/epjc/s10052-014-2762-1
  [arXiv:1311.0048 [hep-ph]].
  %%CITATION = doi:10.1140/epjc/s10052-014-2762-1;%%
  %28 citations counted in INSPIRE as of 03 May 2017
 
  %\cite{KunnawalkamElayavalli:2017hxo}
\bibitem{KunnawalkamElayavalli:2017hxo}
  R.~Kunnawalkam Elayavalli and K.~C.~Zapp,
  %``Medium response in \textsc{Jewel} and its impact on jet shape observables in heavy ion collisions,''
  arXiv:1707.01539 [hep-ph].
  %%CITATION = ARXIV:1707.01539;%%
  
  
\bibitem{foot1}
 More precisely, it is possible within \textsc{Jewel} to trade a lower infra-red cut-off of the parton shower for a lower $\alpha_s$  within the tight experimental constraints set by LEP data. This provides some freedom for varying the
amount of radiation versus scattering and therefore recoil. However, this effect has not been explored systematically, it is expected to be small and
we do not discuss it further in the present paper as it will not affect our main conclusions.  
  
 %\cite{Milhano:2015mng}
\bibitem{Milhano:2015mng}
  J.~G.~Milhano and K.~C.~Zapp,
  %``Origins of the di-jet asymmetry in heavy ion collisions,''
  Eur.\ Phys.\ J.\ C {\bf 76} (2016) no.5,  288
  doi:10.1140/epjc/s10052-016-4130-9
  [arXiv:1512.08107 [hep-ph]].
  %%CITATION = doi:10.1140/epjc/s10052-016-4130-9;%%
  %12 citations counted in INSPIRE as of 03 May 2017
 
 %\cite{Zapp:2012ak}
\bibitem{Zapp:2012ak}
  K.~C.~Zapp, F.~Krauss and U.~A.~Wiedemann,
  %``A perturbative framework for jet quenching,''
  JHEP {\bf 1303} (2013) 080
  doi:10.1007/JHEP03(2013)080
  [arXiv:1212.1599 [hep-ph]].
  %%CITATION = doi:10.1007/JHEP03(2013)080;%%
  %75 citations counted in INSPIRE as of 03 May 2017
   
  
  
  
  %\cite{Dasgupta:2013ihk}
\bibitem{Dasgupta:2013ihk}
  M.~Dasgupta, A.~Fregoso, S.~Marzani and G.~P.~Salam,
  %``Towards an understanding of jet substructure,''
  JHEP {\bf 1309} (2013) 029
  doi:10.1007/JHEP09(2013)029
  [arXiv:1307.0007 [hep-ph]].
  %%CITATION = doi:10.1007/JHEP09(2013)029;%%
  %108 citations counted in INSPIRE as of 02 May 2017
  
  %\cite{Larkoski:2014wba}
\bibitem{Larkoski:2014wba}
  A.~J.~Larkoski, S.~Marzani, G.~Soyez and J.~Thaler,
  %``Soft Drop,''
  JHEP {\bf 1405} (2014) 146
  doi:10.1007/JHEP05(2014)146
  [arXiv:1402.2657 [hep-ph]].
  %%CITATION = doi:10.1007/JHEP05(2014)146;%%
  %127 citations counted in INSPIRE as of 02 May 2017
%
%\cite{Cacciari:2008gp}
\bibitem{Cacciari:2008gp}
  M.~Cacciari, G.~P.~Salam and G.~Soyez,
  %``The Anti-k(t) jet clustering algorithm,''
  JHEP {\bf 0804} (2008) 063
  doi:10.1088/1126-6708/2008/04/063
  [arXiv:0802.1189 [hep-ph]].
  %%CITATION = doi:10.1088/1126-6708/2008/04/063;%%
  %4444 citations counted in INSPIRE as of 02 May 2017



%\cite{CMS:2016jys}
\bibitem{CMS:2016jys}
  CMS Collaboration [CMS Collaboration],
  %``Splitting function in pp and PbPb collisions at 5.02 TeV,''
  CMS-PAS-HIN-16-006.
  %%CITATION = CMS-PAS-HIN-16-006;%%
  %14 citations counted in INSPIRE as of 03 May 2017
 
  %\cite{Larkoski:2015lea}
\bibitem{Larkoski:2015lea}
  A.~J.~Larkoski, S.~Marzani and J.~Thaler,
  %``Sudakov Safety in Perturbative QCD,''
  Phys.\ Rev.\ D {\bf 91} (2015) no.11,  111501
  doi:10.1103/PhysRevD.91.111501
  [arXiv:1502.01719 [hep-ph]].
  %%CITATION = doi:10.1103/PhysRevD.91.111501;%%
  %26 citations counted in INSPIRE as of 02 May 2017

%\cite{Chien:2016led}
\bibitem{Chien:2016led}
  Y.~T.~Chien and I.~Vitev,
  %``Probing the hardest branching of jets in heavy ion collisions,''
  arXiv:1608.07283 [hep-ph].
  %%CITATION = ARXIV:1608.07283;%%
  %8 citations counted in INSPIRE as of 03 May 2017

%\cite{Mehtar-Tani:2016aco}
\bibitem{Mehtar-Tani:2016aco}
  Y.~Mehtar-Tani and K.~Tywoniuk,
  %``Groomed jets in heavy-ion collisions: sensitivity to medium-induced bremsstrahlung,''
  JHEP {\bf 1704} (2017) 125
  doi:10.1007/JHEP04(2017)125
  [arXiv:1610.08930 [hep-ph]].
  %%CITATION = doi:10.1007/JHEP04(2017)125;%%
  %4 citations counted in INSPIRE as of 03 May 2017
  
  %\cite{Rajagopal:2016uip}
\bibitem{Rajagopal:2016uip}
  K.~Rajagopal, A.~V.~Sadofyev and W.~van der Schee,
  %``Evolution of the jet opening angle distribution in holographic plasma,''
  Phys.\ Rev.\ Lett.\  {\bf 116} (2016) no.21,  211603
  doi:10.1103/PhysRevLett.116.211603
  [arXiv:1602.04187 [nucl-th]].
  %%CITATION = doi:10.1103/PhysRevLett.116.211603;%%
  %5 citations counted in INSPIRE as of 10 Jul 2017

  %\cite{Casalderrey-Solana:2016jvj}
\bibitem{Casalderrey-Solana:2016jvj}
  J.~Casalderrey-Solana, D.~Gulhan, G.~Milhano, D.~Pablos and K.~Rajagopal,
  %``Angular Structure of Jet Quenching Within a Hybrid Strong/Weak Coupling Model,''
  JHEP {\bf 1703} (2017) 135
  doi:10.1007/JHEP03(2017)135
  [arXiv:1609.05842 [hep-ph]].
  %%CITATION = doi:10.1007/JHEP03(2017)135;%%
  %14 citations counted in INSPIRE as of 10 Jul 2017
 
 %\cite{Tachibana:2017syd}
\bibitem{Tachibana:2017syd}
  Y.~Tachibana, N.~B.~Chang and G.~Y.~Qin,
  %``Full jet in quark-gluon plasma with hydrodynamic medium response,''
  Phys.\ Rev.\ C {\bf 95} (2017) no.4,  044909
  doi:10.1103/PhysRevC.95.044909
  [arXiv:1701.07951 [nucl-th]].
  %%CITATION = doi:10.1103/PhysRevC.95.044909;%%
  %6 citations counted in INSPIRE as of 10 Jul 2017
 
  %\cite{Chen:2017zte}
\bibitem{Chen:2017zte}
  W.~Chen, S.~Cao, T.~Luo, L.~G.~Pang and X.~N.~Wang,
  %``Evidence of jet-induced medium excitation in $\gamma$-hadron correlation in A+A collisions,''
  arXiv:1704.03648 [nucl-th].
  %%CITATION = ARXIV:1704.03648;%%
  %2 citations counted in INSPIRE as of 13 Jul 2017
  
  %\cite{Wang:2016fds}
\bibitem{Wang:2016fds}
  X.~N.~Wang, S.~Y.~Wei and H.~Z.~Zhang,
  %``Effect of medium recoil and $p_T$ broadening on single inclusive jet suppression in high-energy heavy-ion collisions,''
  arXiv:1611.07211 [hep-ph].
  %%CITATION = ARXIV:1611.07211;%%
  %5 citations counted in INSPIRE as of 13 Jul 2017
  
 
 
 

\end{thebibliography}

\end{document}